# Random Sequential Adsorption on Imprecise Lattice

Vladimir Privman*  and  Han Yan

Department of Physics, Clarkson University, Potsdam, NY 13699, USA

______________________________________________

*Corresponding author. E-mail: privman@clarkson.edu

**ABSTRACT**

We report a surprising result, established by numerical simulations and analytical arguments for a one-dimensional lattice model of random sequential adsorption, that even an arbitrarily small imprecision in the lattice-site localization changes the convergence to jamming from fast, exponential, to slow, power-law, with, for some parameter values, a discontinuous jump in the jamming coverage value. This finding has implications for irreversible deposition on patterned substrates with pre-made landing sites for particle attachment. We also consider a general problem of the particle (depositing object) size not an exact multiple of the lattice spacing, and the lattice sites themselves imprecise, broadened into allowed-deposition intervals. Regions of exponential vs. power-law convergence to jamming are identified, and certain conclusions regarding the jamming coverage are argued for analytically and confirmed numerically.

**J. Chem. Phys., in print (2016)**

## 1. INTRODUCTION

The model of random sequential adsorption (RSA), reviewed in Refs. 1-5, has been widely used to describe processes of irreversible monolayer deposition on various substrates. It applies in situations when time-scales of on-substrate relaxation of the deposited objects are large enough to assume that the attachment is practically irreversible. The attached objects block surface area or substrate sites that they cover. Therefore deposition attempts due to a continuing flux to the surface of the later-arriving objects that overlap those already present are rejected. Ultimately, for large times a jammed-state monolayer deposit is formed with no voids left large enough for new deposition events. Numerical studies of many one-dimensional (1D) and higher-dimensional (especially two-dimensional, 2D) model variants have been reported,[1-5] with emphasis on the properties of and approach to the jammed deposit state. In some cases exact-solution results[1,6] (mostly for 1D) and asymptotic large-time behavior[4,7,8] have been obtained analytically.

Traditionally, two RSA model variants have been studied extensively.[1-6] For deposition on continuum substrates for large times the remaining gaps for adding new objects can be arbitrarily "tight" and are therefore reached with very small probability for a uniform flux of the arriving objects that attempt deposition. As a result, analytical arguments[4,7,8] and exact 1D results[6] supported by numerical evidence[1-5] suggest that approach to the jammed state is slow, power law. For lattice substrates with fixed object arrival rate per lattice site, approach to the jammed state is instead fast, exponential.[1-6] Crossover from lattice to continuum behavior as the lattice spacing becomes very small as compared to the object size has also been studied.[4,6]

With the advent of nanotechnology, not only the depositing objects but also *target substrates* can be tailored to control the resulting structures. Generally, growth and deposition processes can be experimentally realized on patterned 1D substrates made as lines,[9-14] nanotubes,[15-17] etc., or various types of 2D patterned surfaces.[18-33] Specifically, pre-patterned substrates have been studied in several experiments aiming at applications ranging from electronics[15-19] to photovoltaics/optics/optoelectronics,[20-23] to sensors/microarrays,[24-29] and to directed crystal growth/particle assembly.[9-14,30] Here we focus on some interesting properties of and report surprising 1D-modeling findings for RSA on imprecisely prepared substrates. Consideration of RSA on substrates that have various type of randomness has been reported, for instance, in Refs. 34-40. Recently, motivated by nano-patterning, modeling of RSA on substrates has been initiated[5,41] that have the lattice arrangement of the "landing regions" for particle-center deposition, but, due to the details of the particle and landing-region geometries do not enforce a precise lattice alignment of the deposited particles.

New experimental capabilities[9-33] to pre-pattern substrates with preferential sites for specific particle attachment can be viewed as a potential mechanism to form denser deposits fast, with coverage exponentially converging to the jammed state value. In this work we focus on this expectation — which seems natural because the surface is not continuous — and we report a surprising finding that it might actually be not generally correct. Even an infinitesimal "broadening" of the sites of a pre-patterned substrate, deviating from the mathematically precise lattice-point structure might cause the deposition process to revert to power-law large-time behavior, despite the fact that the landing regions are in a lattice arrangement. The specifics depend on the system details. Here we consider a rather general model of 1D deposition of segments of length $a$, on a lattice of spacing $\ell$ between its sites, which instead of just being lattice points are symmetrically broadened (about the lattice points) into segments of width $w$ in which the centers on the depositing objects can land. The precise lattice deposition is then obtained for $w = 0$. For example, in the latter case ($w \equiv 0$) the choice

$$a = n\ell \tag{1}$$

corresponds to the standard lattice $n$-mer deposition ($n = 1, 2, ...$) with exponential convergence to the jammed state. Surprisingly, for arbitrarily small $w > 0$ we find that the deposition becomes exactly the same as continuum deposition with power-law convergence to the jammed state.

The rest of the article is organized as follows. Section 2 introduces the model and presents initial numerical results for dimer RSA on an imprecise substrate vs. continuous "car parking model" (i.e., RSA of segments on a line), alluding to our main conclusion. Section 3 offers analytical arguments for the validity of our main result, which is that, even for an infinitesimal imprecision in the lattice site localization, $n$-mer RSA discontinuously changes to car-parking. In Sec. 4, we consider a more general case of the object size not an exact multiple of the lattice spacing. We offer concluding remarks in Sec. 5.

## 2. FROM CONTINUOUS TO DIMER DEPOSITION

Before addressing the possibility of an imprecisely prepared 1D lattice substrate, we briefly summarize some results for the two illustrative "standard" cases: the continuum 1D car-parking RSA and exact-lattice 1D RSA of dimers, in the notation and context useful for our later discussion. The continuum-substrate 1D RSA, also called the car parking model considers a flux, $\Phi$, of objects to the linear substrate. Arriving objects can attach on contact, anywhere that they do not overlap already deposited objects. In our case it is convenient to also define the rate at which objects arrive per each interval of length $\ell$ of the substrate,

$$R = \Phi\ell, \tag{2}$$

which will be the rate of the deposition attempts per site in the lattice variant of RSA considered later.

The car-parking model is exactly solvable[6] for the density of the centers of the deposited objects, $\rho(t)$, per unit length, as a function of time, $t$. Specifically, the jammed-state fraction of the covered length is

$$a\rho(\infty) \simeq 0.7476, \tag{3}$$

which is known as the Rényi parking constant. We will also cite the result, obtainable from the full solution, that

$$a[\rho(\infty) - \rho(t)] = \frac{e^{-2\gamma}}{2Rt} + \cdots, \tag{4}$$

where the corrections are exponentially small for large time. Here $\gamma \simeq 0.5772$ is Euler-Mascheroni constant. Equation (4) exemplifies the general expectation[4,7,8] that in continuum RSA the maximal coverage is reached according to power-law convergence (in some higher-dimensional cases modified[4,8] with logarithmic-in-time factors).

As an example of the exponential convergence to the jamming coverage in the precise lattice case, let us now briefly summarize results for dimer deposition, $n = 2$ in Eq. (1). Here object centers are exactly aligned with the lattice, obtained by assuming that the attachment attempts, at the rate $R$ per lattice-spacing interval, see Eq. (2), are no longer spread out but are exactly localized at the lattice sites (points). This exactly solvable process, e.g., Refs. 1, 6, 42, yields the jamming coverage

$$a\rho(\infty) = 1 - e^{-2} \simeq 0.8647, \tag{5}$$

and the exponential convergence to it,

$$a[\rho(\infty) - \rho(t)] = (2e^{-2})e^{-Rt} + \cdots, \tag{6}$$

where the corrections are faster-decaying exponentials.

Let us now consider an *imprecisely-prepared substrate* with each linear-lattice site originally at $x = s\ell$, where $s = 0, \pm 1, \pm 2, ...$, broadened into a narrow interval of width $w$. Thus, we assume that the centers of the arriving dimers can deposit in intervals

$$x \in [s\ell - \tfrac{w}{2}, s\ell + \tfrac{w}{2}]. \tag{7}$$

However, to keep the incoming particle deposition attempt rate the same, we assume that the flux of object centers transported into these intervals is

$$R/w = \Phi\ell/w. \tag{8}$$

The flux towards segments outside the *w*-intervals is zero. Figure 1 illustrates this for a more general case of *a* not necessarily exactly equal $2\ell$ or any other multiple of $\ell$. Obviously, the values of *w* can vary in $[0,\ell]$, and one might conjecture that as *w* increases from 0 towards $\ell$, the deposition process of objects of size $a = 2\ell$ with gradually change from RSA of dimers (for $w = 0$) to the continuous car-parking RSA (for $w = \ell$). However, surprisingly, this is not the case.

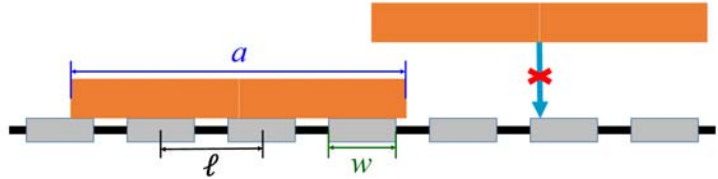

**Figure 1**. An illustration of a deposited object (orange) of size *a*, with its center landed in an interval of width *w* (grey) centered at a site of the 1D lattice of spacing $\ell$. Another object shown, has its center aimed at another allowed interval of width *w*, but its deposition will be rejected because of to its overlap with the already deposited object. Note that here *a* is somewhat larger than $3\ell$.

It is well established that correlations in RSA decay extremely fast,[1,7,8,43,44] i.e., there are no strong fluctuations. Therefore, one can obtain high-precision results numerically on relatively small lattices of size of several hundred $\ell$, without worrying for finite-size effects,[45] and averaging over not too many runs because the system is self-averaging. Such data were obtained for the considered problem in an attempt to explore an expected "crossover" between the continuum and lattice RSA. However, we found a surprising result, later confirmed by an analytical argument (see the next section) that, for any non-zero $w$, no matter how small, the problem discontinuously changes from lattice-RSA (for $w \equiv 0$) to continuum-RSA (for all $0 < w \leq \ell$). Figure 2 illustrates this finding.

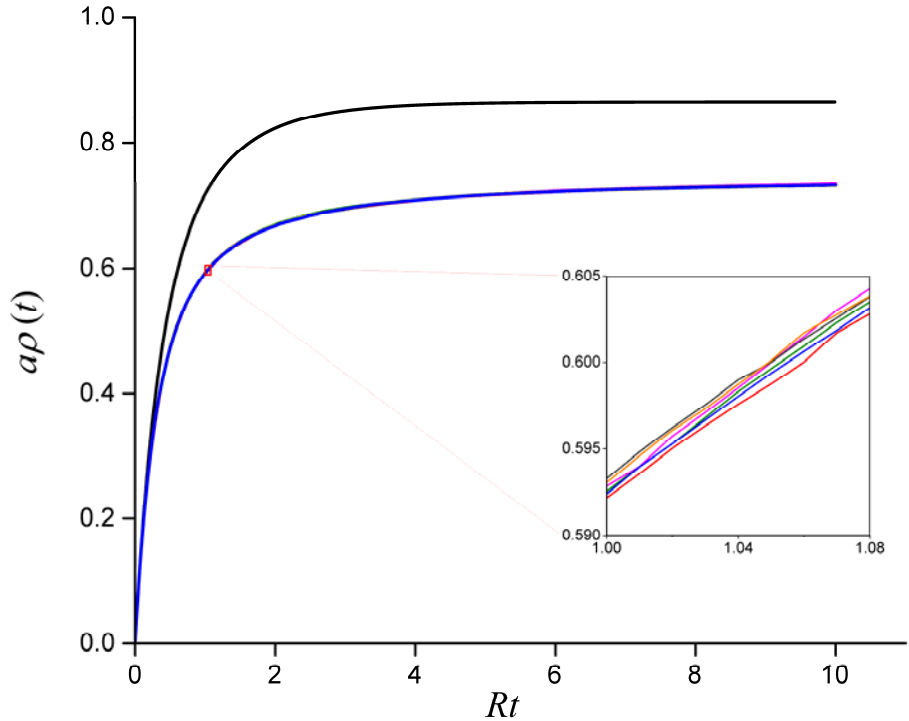

**Figure 2**. Fraction of the covered area for deposition of dimers (the top curve, black, $w = 0$) and deposition on imprecise lattice substrates (data that appears as a single bottom curve) with $w = 0.015\ell$, $0.020\ell$, $0.025\ell$, $0.050\ell$, $0.100\ell$ and $\ell$, where the latter corresponds to continuum car-parking problem. For dimers, we show the exact result, whereas for the six non-zero values of $w$ we show results of numerical Monte Carlo simulations, as described in the text. All the shown $w > 0$ data sets are practically indistinguishable from each other, and also from the the exact solution for $w = \ell$. The latter was not included here, but the six numerically obtained data sets were drawn in different colors (in the order of increasing $w$ values: red, olive, orange, magenta, gray, blue) that, upon magnification of the small red-framed region shown in the Inset, demonstrate some spread due to statistical noise of the Monte Carlo simulation results.

Figure 2 was obtained by a numerical Monte Carlo simulation for a relatively small lattice size, $100\ell$ (and the data were averaged over 5000 runs). With these choices, one can see the spread in the simulation results for the selected $w$ values in the range $0 < w \leq \ell$ due to the statistical noise (the Inset in the figure). Illustrative numerical estimates here as well as in Sec. 4 are accurate to within approximately 1%. Note that the exact $w = \ell$ result is virtually indistinguishable from the shown numerical data for $0 < w \leq \ell$. In the next section, we will argue that, with proper choice of units the full time-dependence of $\rho(t)$ for the infinite-lattice $n$-mer RSA will be the same for all $0 < w \leq \ell$: exactly that of the continuum car-parking problem.

## 3. ANALYTICAL ARGUMENTS FOR *n*-MER DEPOSITION

Let us consider the situation when the arriving objects are exactly $n$-mer with respect to the lattice of spacing $\ell$, i.e., $a = n\ell$. A rejection of a deposition attempt for $w = 0$ can occur due to overlap that can be of length $\ell, 2\ell, \ldots, n\ell$ (between an arriving object that is being blocked by an already deposited object). Broadening the center landing attempt-flux locations to width $w$ introduces another option for rejection of deposition, due to a small overlap of size up to $w$. Other overlaps can also now be somewhat varying in length, by $\pm w$ around each of the values $\ell, 2\ell, \ldots$. We note that the largest overlap, $n\ell$, can only decrease by up to $w$. Thus, the $\pm w$ situation does not occur for $n = 1$ monomers. Blocking of an attempted deposition can of course occur not only on one, as in Fig. 1, but on both ends, but all these considerations are not relevant for the following argument.

We can "rescale" the deposition problem but cutting out lengths in between the allowed-landing intervals of size $w$. In Fig. 1, these are the black segments of size $\ell - w$. The cut-out length can have any value $0 < c \leq \ell - w$, deleted symmetrically from the *middle* of each $\ell - w$ interval. For precise $n$-mer deposition, each of the deposited or arriving objects can be viewed as consisting of $n$ segments of length $\ell$. We will trim each of these segments symmetrically (at both

- 8 -

ends of the ℓ-segment) by $c/2$. As a result of both "cutting" steps, we shrank all the lengths $\ell$ in the problem to $\ell - c \geq w$. However, it is obvious that as long as $c \leq \ell - w$, the rejection (blocking) properties of the depositing objects in the new problem are unchanged. The overlap can have value in the intervals bound by $\ell - c \pm w, 2(\ell - c) \pm w, ...$, etc., and, for $w > 0$, also in the interval of values up to $w$.

It transpires that with the proper redefinitions of the object arrival flux and time scales, the new problem is equivalent to the original one. We note, however, that there is a discontinuity at $w = 0$. The added overlap up to $w$ is not present for $w \equiv 0$, and all the rescaled problems are then equivalent for *n*-mer deposition. Furthermore, as long as $w \equiv 0$, we cannot take the limit $c \to \ell - w$, because all the relevant intervals will then be exactly 0 and the resulting problem will not be well-defined. However, for any *positive* $w < \ell$, we can select $c \equiv \ell - w$, and then *the new substrate will be continuous*. The deposition problem will then be exactly that of car-parking.

Since in the procedure just described object and substrate lengths were rescaled by the same factor $(\ell - c)/\ell$, the fraction of the covered length of the substrate will be unchanged (which defines the "coverage" rescaling) as a function of time, provided we also require that the rate of object arrival (deposition attempt rate) per each interval of length $w$ should also be kept unchanged, equal to $R$. We conclude that *n*-mer deposition on an imprecise substrate with lattice sites broadened to width $w > 0$ is exactly equivalent to car-parking with the flux of objects $\Phi \ell / w = R/w$ to the substrate, with the latter requirement already incorporated in the definition of our model, cf. Eq. (8).

In summary, the added "small overlap" of length up to $w$ between the objects that are involved in the RSA process, with their centers aimed at broadened lattice-site points replaced with landing interval of width $w$, is the key property that causes $w = 0$ *n*-mer deposition to discontinuously turn into car-parking problem for any positive value of $w < \ell$.

## 4. THE GENERAL-*a* CASE

Given the afore-reported interesting observations for $a = n\ell$, one can also consider the more general case of not only the lattice sites being spread to *w*-intervals $0 \leq w \leq \ell$, but also the object size, *a* varied, the latter assuming positive values that are not necessarily multiples of the lattice spacing, $\ell$. (We only offer limited comments below on the deposition of point-like objects, $a = 0$, that have been considered in a different context, e.g., Ref. 47.) Figure 3 depicts the numerically calculated jamming ($t = \infty$) value of the coverage as the number of deposited objects per lattice site, $\ell\rho(t)$, for a range of varying *a* and *w*. It is also instructive to draw the earlier-defined fraction of the covered area, $a\rho(t)$, at jamming, which is shown in Fig. 4 for two different viewing angles.

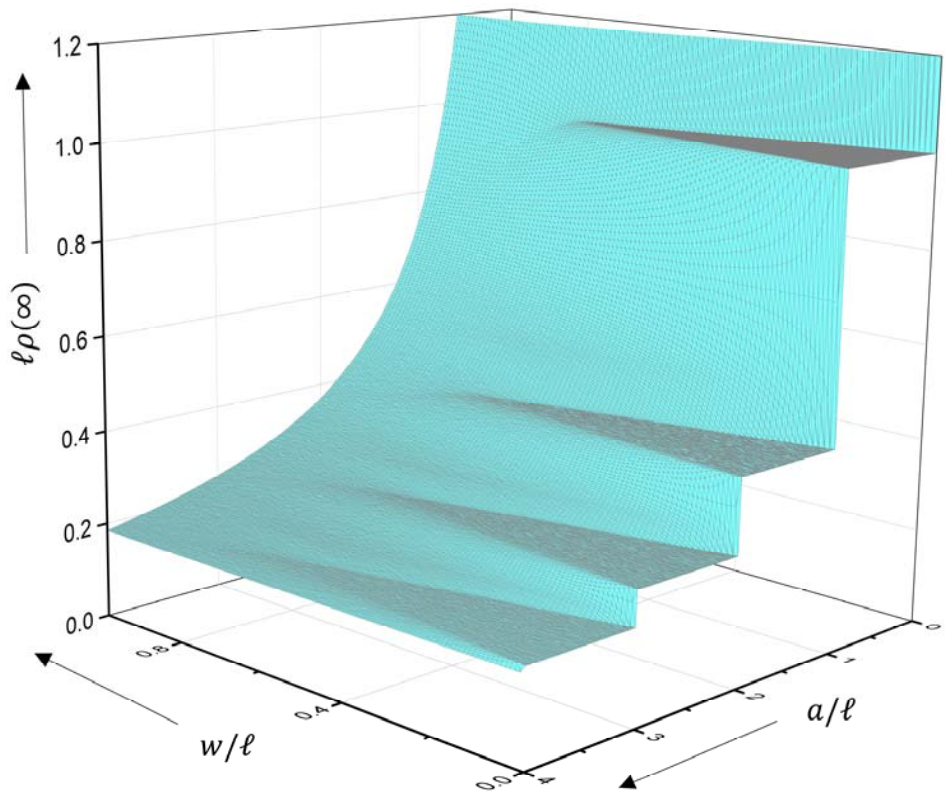

**Figure 3**. Numerically calculated number of deposited objects per lattice site, at jamming.

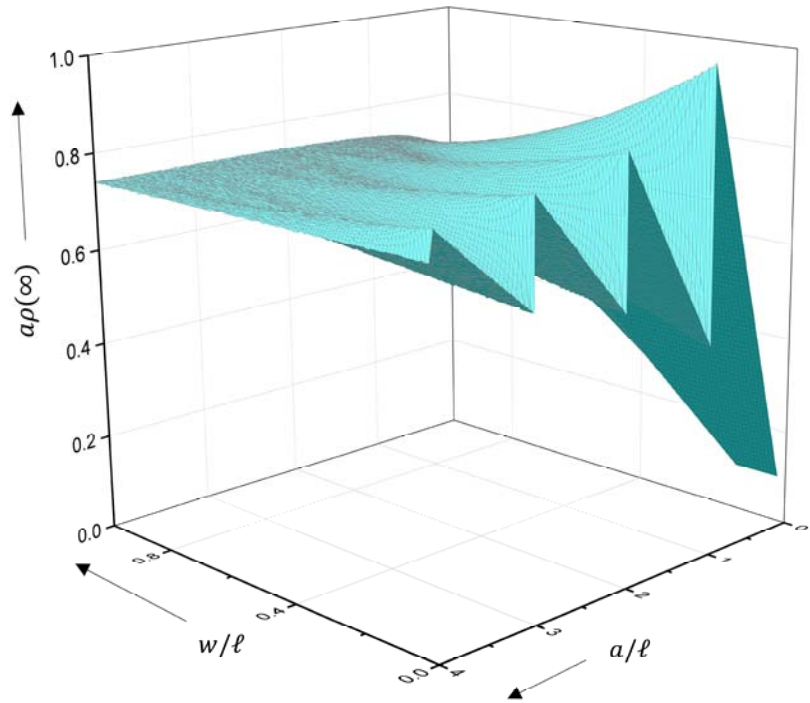

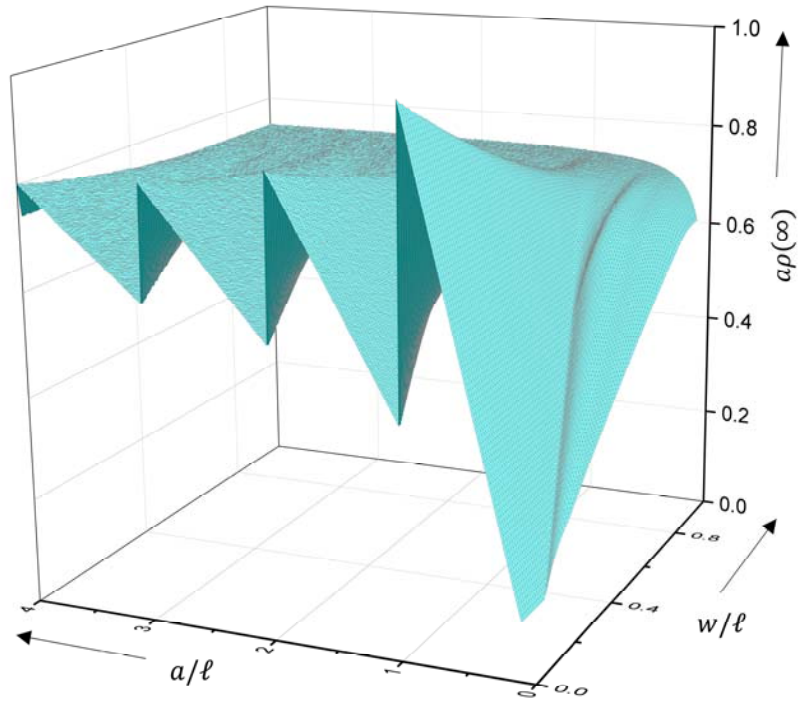

**Figure 4**. The same data as in Fig. 3, but shown as the fraction of the covered area (at jamming), for two different viewing angles.

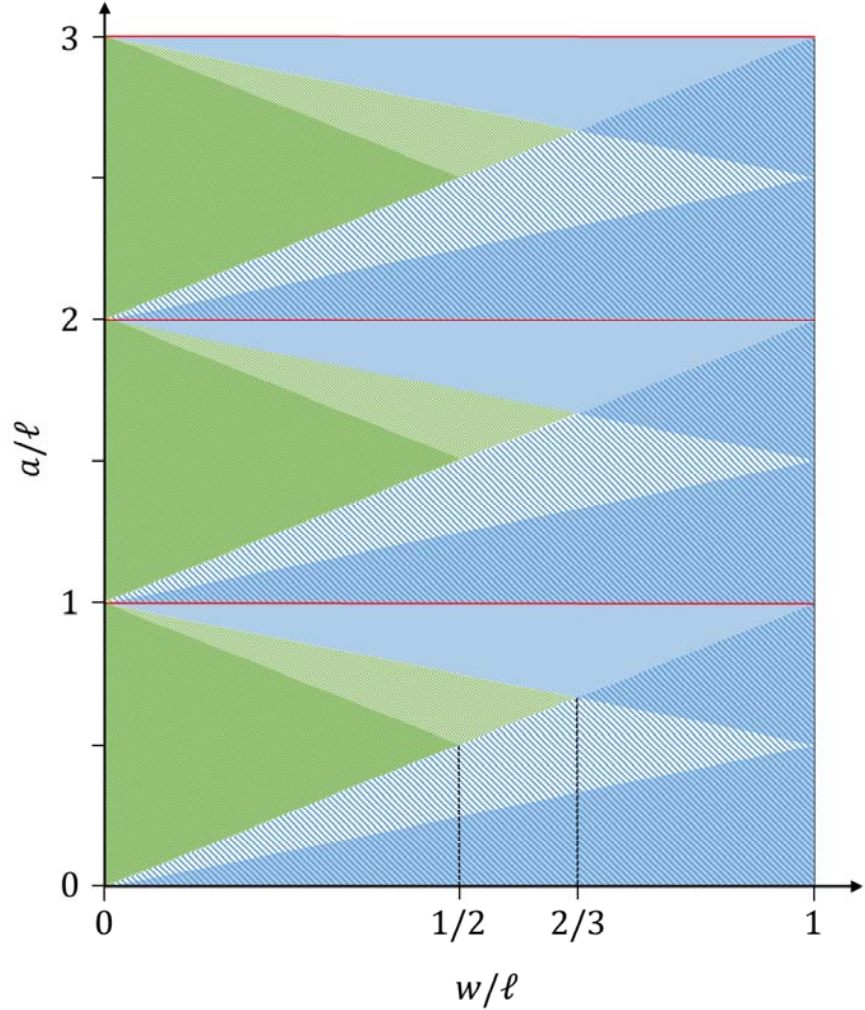

**Figure 5**. The blue-shaded areas are regions of power-law convergence to jamming in the ($w,a$) plane. The solid-blue areas are defined by Eq. (9) and, for the lowest half-area, Eq. (10). The hatched-blue areas are defined by Eq. (11). Exponential convergence to jamming occurs in green-shaded areas, which can also be seen as constant parts of the surface in Fig. 3 and planar parts in Fig. 4. The hatched-green areas are defined by Eq. (12). The solid-green sub-areas are defined by Eq. (13) and correspond to the exact $n$-mer deposition, as described in the text. The red lines correspond to the exact car-parking for any $w > 0$, as discussed in Sec. 2-3.

To map out various region of behavior for varying *w* and *a*, let us offer analytical considerations. We begin by identifying regions for which the approach to jamming is power-law, $\sim 1/t$, because Pomeau argument[7] applies. For this, a distribution of gaps (formed due to earlier-deposited objects) available for arriving objects to deposit into, should include those which are arbitrarily close to the object size (i.e., only infinitesimally larger that *a*). One such configuration is when two objects form such a gap. The distance between their centers is then $2a + \varepsilon$, and therefore for these two centers and also for the center of the arriving (in the middle) object to fall within *w*-intervals, we must have $2k\ell - w < 2a < 2k\ell + w$, which gives

$$k - \frac{w}{2\ell} < \frac{a}{\ell} < k + \frac{w}{2\ell}, \quad k = 1,2,\ldots. \tag{9}$$

Note that the upper half of the possible *a* values can also be realized with $k = 0$, and therefore the approach to jamming will also be power-law for

$$0 < \frac{a}{\ell} < \frac{w}{2\ell}. \tag{10}$$

For the case of Eq. (10), it is important to note that, in some situations the edge of the *w*-interval into which *a*-object centers fit is needed as "the other end" to form small gaps. We do not discuss this in detail here, because the latter mechanism can be used to identify a superseding regime, discussed in the next paragraph. The identified regions are marked in Fig. 5, and for *k* > 0 they are triangular wedges centered at the previously considered (Sec. 2-3) lines $a = k\ell$, with the discontinuous jump to exponential-convergence (to jamming) behavior at the "tip" of each region at *w* = 0.

There is also a possibility that a small gap is formed by an already deposited object having its edge positioned such that an arriving object can only land with its center very close to an edge of a *w*-interval (instead of being blocked by another deposited object on its other side). The distance $a + \varepsilon$ between the centers of the deposited and arriving objects can be realized provided $k\ell < a < k\ell + w$. However, for such small-probability "small-gap" deposition configurations to be statistically relevant, another earlier-deposited (than the attempted arrival under consideration) object must be positioned in such a way that it blocks the "large-gap" deposition of our arriving object in the next *w*-interval, thus only leaving the "small-gap" deposition option. One can check that the condition $k\ell < a$ also ensures the latter requirement. (For $k = 0$ this can be checked separately by simple geometrical considerations.) We therefore conclude that the power-law $\sim 1/t$ convergence also applies in the regions

$$k < \frac{a}{\ell} < k + \frac{w}{\ell}, \qquad k = 0,1,2,\ldots. \tag{11}$$

These regions are also marked in Fig. 5, and they partly overlap with the previously identified regions, Eq. (9) and (10), which means that in those cases more than one small-gap formation mechanism might be possible.

Let us now consider the conditions for which the arriving object can only deposit in gaps that exceed their size, *a*, by at least some fixed length (means, the gaps into which deposition can occur cannot be infinitesimally close to *a*). This implies that the convergence to jamming will be exponential.[4] A set of conditions that gives this behavior is

$$k - \frac{w}{2\ell} > \frac{a}{\ell} > k - 1 + \frac{w}{\ell}, \qquad k = 1,2,\ldots. \tag{12}$$

Here $a > (k-1)\ell + w$ ensures that two objects cannot land with their centers in two *w*-intervals that are $k - 1$ "broadened lattice sites" apart. The second condition is $2k\ell - w > 2a$, which is the reverse of the condition, encountered earlier, that yielded the left-hand side of the inequality in Eq. (9). It ensures that an arriving object that has space to deposit, has a finite (non-infinitesimal) "wiggle space" between its two earlier-deposited neighbors. The regions defined by Eq. (12) are tringles marked in Fig. 5, and they also correspond to the flat regions seen in Fig. 3 (planar regions in Fig. 4). The approach to jamming in them is exponential.

The $w \equiv 0$ sides of the triangles just identified, cover the $w \equiv 0$ axis. As the length of the object, $a$, increases from 0 to $\ell$, then to $2\ell$, etc., the deposition along the $w \equiv 0$ axis discontinuously changes from monomer for $a \leq \ell$ (including for $a = 0$, which corresponds to deposition of pointlike objects in pointlike sites), to dimer for $\ell < a \leq 2\ell$, to trimer for $2\ell < a \leq 3\ell$, …, respectively. However, except for the just described points on the line $w \equiv 0$, the upper boundaries, $a = k\ell - w/2$, with $k = 1,2,...$, of the triangular regions defined by Eq. (12) correspond to continuum approach to jamming, including the "tips" of the triangles at $w/\ell = 2/3$, whereas the lower boundaries (exclusive of their end points at the $w/\ell = 2/3$ "tips"), $a = (k-1)\ell + w$, $k = 1,2,...$, correspond to exponential approach to jamming. We did not explore in detail the nature of possible crossover behaviors as the upper boundaries or their $w/\ell = 2/3$ tips are approached from within these triangular regions, and a possible power-law behavior different from Pomeau's $\sim 1/t$, see Ref. 7, because of potential peculiarities in the small-gap density distribution at these $w/\ell = 2/3$ tips.

The *jamming* density of the object centers per lattice site, $\ell\rho(\infty)$, is the same as for the lattice *n*-mer deposition throughout each region defined by Eq. (12), including at the regions' boundaries, with the same value of *n*, except at the common points on the $w = 0$ axis, at which the lower-*n* region value should be used, as described earlier. This constant-density behavior can be seen in Fig. 3. However, the rate at which the final coverage builds up to its jamming value may vary even when the convergence is exponential, because the "wiggle space" left for object-center deposition in parts of *w*-intervals depends on the specific parameter combinations and local configuration. We note that sub-regions in which the actual time dependence of $\ell\rho(t)$ is guaranteed to be exactly *k*-mer, can be identified as follows. We replace Eq. (12) with

$$k - \frac{w}{\ell} > \frac{a}{\ell} > k - 1 + \frac{w}{\ell}, \qquad k = 1,2,... . \tag{13}$$

Here the right-hand side condition $a > (k-1)\ell + w$ is the same as before — discussed in connection with Eq. (12). The second condition, $a < k\ell - w$, ensures that two depositing objects $k$ lattice positions apart do not block each other no matter where in their respective *w*-intervals their centers land. These sub-regions, which are smaller, symmetrical triangles within the earlier-identified triangular regions, are marked in Fig. 5.

## 5. CONCLUSION

In summary, an interesting conclusion of this study has been that, even a small imprecision in the localization of the lattice landing sites can have dramatic effects on the density of the formed deposit, especially when it is measured as the fraction of the covered substrate area (in 2D; length in 1D). This effect is further amplified if the object size is not a precise multiple of the lattice spacing. Wide swings in the coverage are exemplified in Fig. 4. Furthermore, the convergence to the maximal coverage can also become rather slow, power-law, except in regions of parameter values that are identified to correspond to fast, exponential convergence. Various boundaries that separate regimes of different behavior in the problem parameter space are set by the geometry of exclusion (blocking) of nearby objects, and/or by the geometry of the type considered in asymptotic-convergence (to jamming) arguments.[4,7,8] The latter refer to the possible positioning of more than one earlier-deposited objects to form small gaps into which single additional objects can land, and the distribution of these small-gaps' sizes.[4,7,8]

The reported results suggest that efforts at precise positioning and object-sizing in nano-manufacturing might be counterproductive if done as part of forming structures under practically irreversible "assembly" conditions. A certain degree of relaxation, to allow objects to "wiggle their way" into matching positions may actually be more effective in improving both the density and rate of formation of the desired structures. Added relaxation processes can leave behind certain unresolvable local defects or slow-evolving low-dimensional defect structures (such as defect lines in 2D), e.g., Ref. 48, 49, but they will usually improve the overall degree of order in the bulk structural properties.